\documentstyle[epsfig,wrapfig,longtable]{aipproc}

\newcommand{\eqb}{\begin{eqnarray}}
\newcommand{\eqe}{\end{eqnarray}}
\newcommand{\diff}{{\rm d}}

\begin{document}
\title{Effect of intergalactic absorption in the TeV 
$\gamma$-ray spectrum of Mkn~501}

\author{Alexander K. Konopelko}
\address{Max-Planck-Institut f\"{u}r Kernphysik, \\
Heidelberg D-69029, Postfach 10 39 80, Germany\\
e-mail: alexander.konopelko@mpi-hd.mpg.de}


\maketitle

\begin{abstract}
We discuss an effect of the intergalactic absorption of the TeV 
$\gamma$-rays in time-averaged spectrum of Mkn~501 measured by 
the HEGRA collaboration. Analysis of the spectral behavior, 
variability time scale and relevant calculations of TeV $\gamma$-ray 
emission allow to conclude the presence of a noticeable absorption 
of the TeV $\gamma$-rays in the Mkn~501 energy spectrum.  
\end{abstract}

\section*{Introduction}
The ground-based detectors, utilizing the so-called imaging air 
\v{C}erenkov technique, offer an effective tool to study the cosmic 
TeV $\gamma$-rays. Recently, a number of celestial objects has been 
identified as TeV $\gamma$-ray emitters by use of such technique 
\cite{1}. Among them there are two active galactic nuclei (AGN) -- 
Mkn~421 and Mkn~501 -- which for almost similar redshift of 0.031 and 
0.034 respectively, have very different properties of a TeV $\gamma$-ray 
emission. In particular, TeV $\gamma$-ray fluxes from Mkn~421 and 
Mkn~501 differ in variability time scale and spectral behavior. 
Mkn~421 has shown significant flux variations within a time period as short 
as 15~minutes \cite{2} whereas Mkn~501 may outburst during a period of 
6~months with an extraordinary high $\gamma$-ray flux of more than 
3~Crabs on average \cite{3}. The energy spectrum of Mkn~501, as measured 
in the energy range from 0.5~TeV up to 20~TeV, shows evident curvature 
($dJ_\gamma/dE \propto E^{-1.9} exp(-E/6.2)$) and the spectrum shape does 
not depend on the flux level \cite{4}. At the same time the Mkn~421 energy 
spectrum is very steep and consistent with the pure power law 
($dJ_\gamma/E \propto E^{-3.1}$) over the energy range 0.5-7~TeV, at least 
during low state of emission \cite{5}. All that shows an apparent intrinsic 
difference in the mechanism of the TeV $\gamma$-ray emission which is 
widely believed to be an inverse Compton scattering of electrons within a 
relativistic jet directed along the observer line of site (for review see \cite{1}). 
In addition the measured spectra of TeV $\gamma$-rays from such distant 
sources as Mkn~421 and Mkn~501 might be affected by the $\gamma$-ray 
absorption on the diffuse intergalactic infrared (IR) background. Here we 
discuss how important might be the effect of such absorption on the spectra 
of two observed AGNs in particular Mkn~501 which shows a spectacular shape 
of its spectrum.
\begin{figure}[t]
\begin{center}
\includegraphics[width=0.53\linewidth]{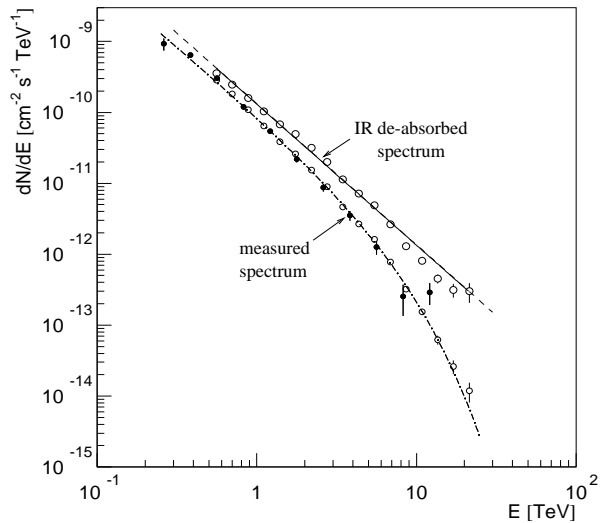}
\end{center}
\caption{The energy spectrum of Mkn~501 as measured by the 
HEGRA IACT array (open circles) \protect\cite{4}). The combined fit (power law 
plus exponent) of the HEGRA data is shown by the dotted-dashed curve. 
The Mkn~501 spectrum measured by the Whipple group (filled circles) is from 
\protect\cite{6}. The ``de-absorbed'' HEGRA data and a power law 
fit (solid line) are shown also.}
\end{figure}    
\section*{Observations}
During an extraordinary outburst of TeV $\gamma$-rays from Mkn~501 in 1997 
observation period this object was monitored by several ground-based imaging 
air \v{C}erenkov telescopes (IACTs) \cite{3}. The HEGRA stereoscopic system of 
4 IACTs has observed Mkn~501 for a total exposure time of $110\,$hours \cite{4}. 
The unprecedented statistics of about 38,000 TeV photons, combined with the good 
energy resolution of $\sim 20$\% allowed determination of a spectrum over the 
energy range from $500\,$~GeV up to $24\,$~TeV. The shape of the spectrum does 
not depend on intensity of the source. It justifies the determination of the 
time-averaged Mkn~501 spectrum. The energy spectrum of Mkn~501, as measured by 
the HEGRA group, shows apparent curvature over entire energy range. The shape 
of the spectrum may be well described by the power law with an exponential 
cutoff. A fit of the data gives:
\eqb
\diff N/\diff E =  10.8\cdot 10^{-11} E^{-1.92}  \exp\left[-E/6.2\right], \,\, 
\rm cm^{-2} s^{-1} TeV^{-1}
\eqe
The detailed systematic analysis of the fit parameters has been discussed in \cite{4}. 
The HEGRA data are also shown in Figure~1.

\begin{figure}[t]
\begin{center}
\includegraphics[width=0.53\linewidth]{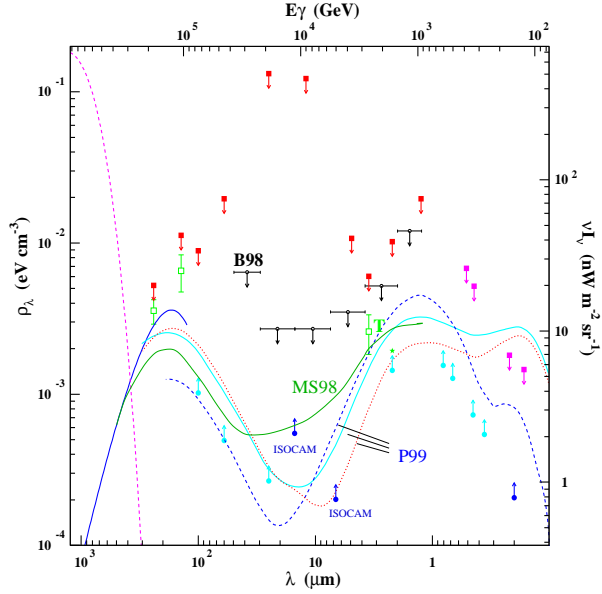}
\end{center}
\caption{Compilation of data and models for a spectral energy distribution of the diffuse 
intergalactic background taken from \protect \cite{1} (adapted). The results of calculations 
using the model \protect \cite{14} for high IR photon field are shown by solid curve 
and denoted by MS99. Calculations from \protect \cite{15} for different model parameters 
are shown by dotted, dashed, thin solid curves and denoted by P99.} 
\end{figure}

\section*{Discussion on spectrum shape}

The curvature in Mkn~501 energy spectrum may be caused by several reasons. The 
curved energy spectrum of TeV $\gamma$-rays may be attributed to (i) the intrinsic 
spectrum of TeV $\gamma$-ray emission within the synchrotron self-Compton or 
external inverse Compton scenarios; (ii) the curvature might be due to the 
absorption of TeV $\gamma$-rays by the pair production inside the source, or (iii) 
in intergalactic medium; finally, the observed energy spectrum may be affected 
by a combination of several reasons noticed above. 

The recent calculations based on the synchrotron self-Compton (SSC) and external 
Compton (EC) models could explain rather well currently established variability 
time scales of X-ray and TeV emission of Mkn~421 and Mkn~501 (see e.g., \cite{7}). 
Thus the observation of the variability of TeV $\gamma$-ray flux at the level of 
$\simeq$1~hr \cite{8}, at least, limits the Doppler boosting factor of the emitting 
jet as $\delta \geq 10$. For such big Doppler boosting factor the $\gamma$-ray absorption 
within the sources does not play an important role and $\gamma$-ray photons can easily 
escape from the emitting region \cite{9}. 

The shape of Mkn~501 energy spectrum as measured by the HEGRA collaboration can 
not be easily fitted by pure SSC and EC models (see e.g., \cite{9}). In particular, 
the shape of a spectrum appears to be very steep above 5~TeV. In addition, the 
observations with Rossi X-Ray Timing Explorer have shown that the spectrum of Mkn~501 
varies strongly with generally a very hard spectral index extending to much higher 
energies ($\geq 100$~keV) \cite{10}. On the contrary, the TeV $\gamma$-ray spectrum 
does not show any variations in the spectrum shape \cite{4}. The simultaneous variations 
in the X-ray and TeV $\gamma$-ray fluxes may be well described by a change in the maximum 
energy to which electrons can be accelerated $\gamma_{max}$ \cite{11} (hereafter 
$\gamma_{max}$ is a corresponding maximum Lorentz factor). Thus the energy spectrum of 
TeV $\gamma$-rays can be extremely soft, e.g., $\alpha \geq 3.0$ ($\alpha$ is 
an index of a power law energy spectrum), due to the cutoff in the spectrum of emitting 
electrons. However for that the variations in $\gamma_{max}$ lead to significant change 
of the spectrum slope in TeV $\gamma$-rays which is not a case for the HEGRA observations. 
It is more likely that synchrotron photons of approximately 1-20~keV, emitted by the 
electrons accelerated within the jet, are up scattered by the same electrons to the TeV 
energies. For such scenario the X-ray variabilities caused by the hardening of the 
initial electron spectrum not necessarily lead to a variations of the spectrum slope in 
TeV $\gamma$-rays which is relatively flat $\alpha \simeq 2.0$ and remains constant. 
Interestingly, the TeV energy $\gamma$-ray spectrum as measured by HEGRA shows very 
similar spectrum slope in the energy range below 5~TeV whereas it deviates from the 
power law strongly in the high energy part. Such behavior might be easily explained by 
the effect of intergalactic absorption. 

\section*{IR absorption in TeV spectrum of Mkn~501}

While propagating in the interstellar medium the TeV $\gamma$-rays may attenuate through 
pair production process in the intergalactic infrared radiation field (IIRF) \cite{12}. 
The corresponding opacity of intergalactic medium is determined by the 
spectral energy distribution (SED) of the IR photon field (see Figure~1). The absorption 
of $\gamma$-rays in the energy range from 0.5 to 20~TeV rely on IR SED of photon field 
between 1 to 50~mkm. Recently measurements as well as low upper limits of SED strongly 
constrain the shape of SED in the range relevant to the TeV $\gamma$-rays. Compilation 
of present data is shown in Figure~1. We also show two models of SED from \cite{13} and 
\cite{14}. Note that the recent tentative detection of IR photons at 3.5~mkm by COBE 
\cite{15} is consistent with both models whereas the ISOCAM lower limit 
on IR photon field, if true, favour model from \cite{13} with rather flat SED at mid IR region. 
An optical depth of $\gamma$-ray absorption as a function of energy and redshift, 
$\tau =\tau (E_\gamma, z)$, was calculated in \cite{16} using the predictions on SED of 
intergalactic IR photon field according to \cite{13}. As such these data may be used to unfold 
the Mkn~501 energy spectrum measured by the HEGRA group, $(dN_\gamma/dE)_{m}$, in order to 
get a ``de-absorbed'' intrinsic energy spectrum of Mkn~501, $(dN_\gamma/dE)_{i}$.
\begin{equation}
(dN_\gamma/dE)_{i}=(dN_\gamma/dE)_m \cdot e^{\tau (E,z)}
\end{equation}
The ``de-absorbed'' HEGRA data are shown in Figure~1 together with a power law fit. 
We find \cite{17} that the data points can be well fitted by 
\begin{equation}
(dN_\gamma/dE)_i = (1.32\pm 0.04) \cdot 10^{-10} (E/1~TeV)^{-2.0\pm 0.03}.
\end{equation}
Note that similar results have been shown at this Workshop by the Telescope Array group 
using their measurement of the Mkn~501 TeV $\gamma$-ray spectrum. We show in Figure~3 
the large scale spectral energy distribution of Mkn~501 calculated assuming the 
absorption. 

\section*{Comparison of Mkn~501 and Mkn~421 spectra}

\begin{figure}
\begin{center}
\includegraphics[width=0.5\linewidth]{f-mkn501.epsi} \\
\end{center}
\caption{Spectral energy distribution of Mkn~501 computed using the homogeneous 
model \protect \cite{11}. Calculations have been done assuming the 
variability time scale of $t_{var} = 10^4$s, maximum Lorentz factor of the emitting 
electrons: $\gamma_{max} = 1.4\cdot 10^6$, magnetic field: $B=0.7\cdot 10^{-3}$G, 
Doppler boosting factor: $\delta = 80$. The HEGRA data are from \protect \cite{4}.}
\begin{center}
\includegraphics[width=0.5\linewidth]{f-mkn421.epsi} 
\end{center}
\caption{Spectral energy distribution of Mkn~421 calculated in \protect \cite{11}          
($t_{var} = 10^3$s, $\gamma_{max}=2\cdot 10^5$, $B=2.5\cdot 10^{-3}$G, $\delta =47$). 
The HEGRA data are from \protect \cite{5}.}
\end{figure}
As reported in \cite{18} the spectra of Mkn~421 and Mkn~501, measured by the 
Whipple group in the high state of emission, show noticeable difference in their spectral shape 
over the energy range 0.3-10~TeV. The spectrum of Mkn~421 is a power law whereas the 
spectrum of Mkn~501 is apparently curved. Since two objects Mkn~421 and Mkn~501 have almost the 
same red shift one may conclude that these two objects have 
different intrinsic energy spectra of TeV $\gamma$-rays \cite{18}. However the 
Whipple data for the Mkn~421 and Mkn~501 energy spectra at the energies above 1~TeV do not show 
prominent difference and both could be well fitted by power law. Apparent difference in two 
spectra is at energies less than 1~TeV, namely in the range where the absorption of TeV 
$\gamma$-rays in the intergalactic IR photon field does strongly affect the spectra. 
Similar behavior of both spectra in the 
energy range above 5~TeV does not contradict the effect of absorption at these energies as stated 
above. 
The spectrum of Mkn~421 as measured by HEGRA collaboration in low state shows power law behavior 
$dN_\gamma/dE \propto E^{-3.1}$ \cite{5}. The HEGRA data allow to extend the spectral 
measurements only up to 7~TeV. Such steep spectrum most likely can be attributed to the very soft 
intrinsic source spectrum and may not disprove the effect of absorption of TeV $\gamma$-rays 
(see Figure~3).

\section*{Conclusion}

We propose a possible scenario explaining the spectral shape of the Mkn~501 energy spectrum 
as measured by HEGRA collaboration. Strong variations of X-ray emission argue in favor of 
rather flat intrinsic spectrum of TeV $\gamma$-rays. We conclude that absorption in the 
interstellar IR photon field plays an important role and produces an apparent curvature 
observed in Mkn~501 spectrum. The SSC fit of the spectral shape constrain rather high value 
of the Doppler boosting factor of a emitting jet, $\delta>50$. Future multi-wavelength 
observations as well as detections of other BL Lac objects will help in future understanding 
of mechanisms of the TeV $\gamma$-ray emission and propagation processes.

\end{document}